\documentclass[conference,9pt]{IEEEtran}
\IEEEoverridecommandlockouts
\usepackage{cite}
\usepackage{url}
\usepackage{amsmath,amssymb,amsfonts}
\usepackage{algorithmic}
\usepackage{graphicx}
\usepackage{textcomp}
\usepackage{xcolor}
\usepackage{array}
\usepackage[switch]{lineno}

\newcommand{\ours}{\textit{MixMoMo}}
\def\BibTeX{{\rm B\kern-.05em{\sc i\kern-.025em b}\kern-.08em
    T\kern-.1667em\lower.7ex\hbox{E}\kern-.125emX}}

\renewcommand{\footnoterule}{%
    \kern -3pt
    \hrule width 0.4\columnwidth height 0.4pt
    \kern 2.6pt
}

\begin{document}

\title{MixMoMo: Accelerating Extra--Large MoE Inference with Mixed--Precision GPU--NPU Execution on Mobile Devices}

\author{
    \IEEEauthorblockN{
        Hailong Zou\textsuperscript{1,*},
        Maoliang Li\textsuperscript{1,*},
        Taohong Han\textsuperscript{3},
        Haoze Chi\textsuperscript{2},
        Zihao Zheng\textsuperscript{1},
        Jiayu Chen\textsuperscript{1},
        Xiang Chen\textsuperscript{1,\ensuremath{\dagger}}
    }
    \IEEEauthorblockA{
        \textsuperscript{1}School of Computer Science,
        Peking University, Beijing, China\\
        \textsuperscript{2}School of Electronics Engineering and Computer Science,
        Peking University, Beijing, China\\
        \textsuperscript{3}School of Artificial Intelligence,
        Tianjin University, Tianjin, China
    }
}

\maketitle

\begingroup
\renewcommand\thefootnote{}
\footnotetext{\textsuperscript{*} Both authors contributed equally to this research.}
\footnotetext{\textsuperscript{\ensuremath{\dagger}} Corresponding author: Xiang Chen, xiang.chen@pku.edu.cn.}
\endgroup

\begin{abstract}
Mobile agents and on-device intelligence increasingly demand more capable models on smartphones.
Through sparse expert activation, Mixture-of-Experts (MoE) scales model capacity, yet growing numbers of fine-grained experts complicate loading and computation.
Model compression complements expert offloading by reducing storage and computation costs.
Among compression techniques, mixed-precision quantization assigns lower expert bitwidths to balance model quality and inference efficiency.
However, existing offline metrics and routing scores fail to reflect expert quantization sensitivity, undermining precision assignment.
Diverse bitwidths also complicate balanced expert allocation across GPUs and NPUs with distinct execution characteristics, while tile-wise dequantization underutilizes NPU vector parallelism.
We present \ours{}, an efficient mixed-precision inference runtime for extra-large MoE models on heterogeneous mobile SoCs.
\ours{} combines offline quantization-impact profiles with activation-based runtime risk estimates for quality-constrained expert bitwidth assignment.
It further adopts precision-structured execution for balanced GPU--NPU expert allocation and HVX-parallel dequantization for efficient NPU vector utilization.
Across three MoE models and five benchmarks on two mobile platforms, \ours{} achieves average decoding speedups of 3.72$\times$ over uniform Q4\_0 and 1.53$\times$ over the strongest mixed-precision MoE baseline.
\end{abstract}


\section{Introduction}
Advances in mobile agents and on-device intelligence drive demand for deploying more capable models on smartphones~\cite{wang2024mobileagentv2,xue2024powerinfer2,xu2025edgellm}.
Extra-large Mixture-of-Experts (MoE) models expand model capacity by scaling up expert parameters, but massive parameter sizes make deployment on smartphones challenging~\cite{shazeer2017moe,fedus2022switch,yi2025edgemoe}.
Meanwhile, relatively uniform expert activation distributions across fine-grained experts in extra-large MoE models further complicate expert loading and execution~\cite{yang2025qwen3,deepseekai2024deepseekv3,yi2025edgemoe}.
Orthogonal to prior work that enables extra-large MoE inference on smartphones through expert offloading~\cite{tang2024hobbit,li2026bigmomo}, model compression directly reduces model storage and computation costs to facilitate deployment.
Among compression techniques, mixed-precision quantization assigns different bitwidths to experts to balance model quality and inference efficiency~\cite{huang2025mcmoe,duanmu2025mxmoe}.

Modern smartphones integrate heterogeneous computing engines, including CPUs, GPUs, and NPUs, to enhance computing performance~\cite{qualcomm2025snapdragon8gen5}.
Among them, GPUs and NPUs are better suited than CPUs for accelerating MoE expert matrix operations, with GPUs offering flexible parallel computation and NPUs leveraging specialized matrix engines~\cite{hao2026scaling}.
However, disparities in execution performance and cross-device synchronization overhead make it difficult to fully exploit the computational resources of both GPUs and NPUs~\cite{chen2025heteroinfer,li2026hero,chen2026ditxpu}.
These architectural differences also affect low-bitwidth format support and quantized expert execution efficiency.

Given these characteristics, efficient mixed-precision inference of extra-large MoE models on heterogeneous mobile SoCs faces three interrelated challenges.
\textit{\textbf{Precision Assignment}} determines expert bitwidths based on estimated expert importance.
Expert importance can be characterized by various metrics, including activation frequency, quantization error, and routing scores~\cite{huang2025mcmoe,duanmu2025mxmoe}.
\textit{\textbf{Expert Allocation}} then distributes activated experts across GPUs and NPUs based on workload characteristics~\cite{duanmu2025mxmoe,chen2025heteroinfer}.
After allocation, \textit{\textbf{Quantized Execution}} processes low-bitwidth experts through different weight preparation and matrix computation flows on GPUs and NPUs~\cite{hao2026scaling}.

\begin{figure}[t]
    \centering
    \includegraphics[width=\columnwidth]{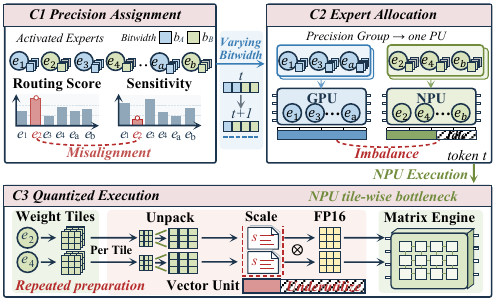}
    \caption{Challenges in Mixed-Precision Inference of Extra-Large MoE.}
    \label{fig:system_overview}
    \vspace{-4mm}
\end{figure}

For \textit{precision assignment}, existing offline methods estimate expert importance using activation frequency and quantization error~\cite{huang2025mcmoe,duanmu2025mxmoe}, whereas more recent online methods rely on routing scores, the per-token gating weights for expert selection and output aggregation~\cite{tang2026moeapex,kim2026dpamoe}.
However, relatively uniform expert activation distributions and dynamic precision requirements undermine offline assignments~\cite{deepseekai2024deepseekv3,wang2024lossfree,chu2025dynaexq}.
Meanwhile, routing scores do not accurately reflect the extent of quantization-induced output deviations, which is defined as \textit{quantization sensitivity}.
Consequently, \textbf{\textit{the misalignment between routing scores and expert quantization sensitivity}} can lead to suboptimal precision assignment.
For \textit{expert allocation}, dynamic precision assignment produces shifting compositions of different-bitwidth experts across tokens.
Consequently, \textbf{\textit{the interplay between varying precision compositions and GPU--NPU execution characteristics}} makes it difficult to allocate experts across GPUs and NPUs to minimize latency.
For \textit{quantized execution}, GPUs fuse low-bit weight preparation with matrix computation, whereas NPUs require weights in hardware-compatible formats~\cite{hao2026scaling}.
To meet these format requirements, NPUs prepare each low-bit expert weight tile through scale metadata preparation, weight unpacking, and FP16 reconstruction.
Consequently, \textbf{\textit{independent tile-wise processing}} incurs repeated preparation overhead and underutilizes NPU vector parallelism.

To address these challenges, we propose \ours{}, an efficient mixed-precision inference runtime for extra-large MoE models on heterogeneous mobile SoCs.
We reveal that \textit{activation magnitudes in critical MoE input dimensions exhibit a strong correlation with expert quantization sensitivity}.
Building on this insight, \ours{} combines offline quantization-impact profiles with runtime activation-based risk estimates to assign expert bitwidths under a quality budget.
After precision assignment, \ours{} dynamically allocates routed-expert groups across GPU and NPU based on device-specific execution-time estimates to balance workloads and reduce latency.
For experts allocated to the NPU, \ours{} batches compatible dequantization tiles across same-precision experts within each computation stage to improve NPU throughput and reduce execution latency.

Evaluations on two OnePlus smartphones across three extra-large MoE models and five benchmarks demonstrate that \ours{} maintains near-FP16 model quality while achieving average decoding speedups of 3.72$\times$ over uniform Q4\_0 and 1.53$\times$ over the strongest mixed-precision MoE baseline for each evaluated model.

\section{Background}

\subsection{Extra-Large MoE Deployment}

\begin{figure}[t]
    \centering
    \includegraphics[width=\columnwidth]{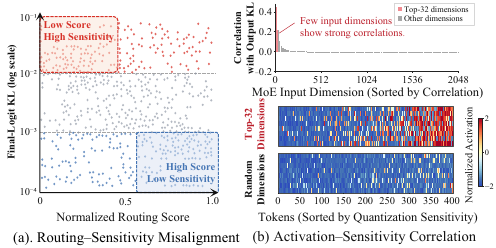}
    \caption{Expert quantization sensitivity.}
    \label{fig:expert_precision}
    \vspace{-4mm}
\end{figure}

Extra-large MoE models have massive parameter counts, with expert weight footprints often exceeding smartphone DRAM capacity.
Due to limited DRAM capacity, expert offloading stores expert weights in flash and loads activated experts into DRAM on demand.
EdgeMoE~\cite{yi2025edgemoe} predicts expert activation and preloads expert weights before computation.
Cache-Conditional Experts~\cite{skliar2025cacheconditional} improves expert cache locality through cache-aware routing.
BigMoMo~\cite{li2026bigmomo} leverages speculative decoding to decouple expert loading and computation, enabling 80B-scale MoE inference on mobile SoCs.

Orthogonal to expert offloading, model compression reduces model storage and computation costs.
Among compression techniques, quantization reduces expert weight size and arithmetic costs through lower-bitwidth representations~\cite{lin2024awq}.
Mixed-precision quantization assigns different bitwidths to individual experts based on their estimated importance, preserving higher precision for more important experts~\cite{huang2025mcmoe,duanmu2025mxmoe}.
It can also be integrated with expert offloading to better support extra-large MoE deployment on smartphones.

\subsection{Expert Precision Assignment}
Expert precision assignment has evolved from static offline configurations toward runtime adaptation~\cite{zheng2026dynamo}.
For offline precision assignment, MC-MoE~\cite{huang2025mcmoe} assigns expert bitwidths based on activation frequency and quantization error, while MxMoE~\cite{duanmu2025mxmoe} combines hardware-aware precision assignment with mixed-precision GroupGEMM.
More recent methods adapt expert precision at runtime using routing information.
MoE-APEX~\cite{tang2026moeapex} selects cache-miss expert precision based on routing scores, while DPA-MoE~\cite{kim2026dpamoe} combines routing scores with offline reconstruction residuals.

Extra-large MoE models contain numerous fine-grained experts but activate a small subset per token~\cite{yang2025qwen3,deepseekai2024deepseekv3}.
Relatively uniform expert activation distributions weaken activation-frequency-based importance estimates, while offline assignments cannot adapt bitwidths to dynamic precision requirements.
Meanwhile, the expanded expert pool increases the complexity of runtime precision assignment.

\subsection{Heterogeneous Expert Execution}
Mobile GPUs and NPUs employ distinct computing units and hardware-specific execution flows for low-bit expert computation.

\textbf{\textit{GPU-Side:}}
Mobile GPUs employ programmable shader processors (SPs) for parallel computation.
For low-bit expert execution during decoding, SPs fuse weight unpacking, dequantization, and matrix--vector multiplication (GEMV) within a unified kernel~\cite{lin2024awq}.

\textbf{\textit{NPU-Side:}}
In contrast, Snapdragon Hexagon NPUs partition quantized expert weights into tiles and load them with quantization scales from DRAM into Vector Tightly Coupled Memory (VTCM).
Hexagon Vector eXtensions (HVX) then unpacks and dequantizes each tile independently into FP16 before matrix computation~\cite{qualcomm2026hmx}.
These operations introduce non-negligible preparation overhead.

Different expert bitwidths involve different compressed weight sizes and distinct unpacking/scaling requirements.
Combined with the distinct weight preparation and matrix computation flows on GPUs and NPUs, these bitwidth-dependent requirements lead to differences in expert execution costs across bitwidths and devices.

\section{Motivation}

\subsection{Expert Quantization Sensitivity Characterization}

\begin{figure}[t]
    \centering
    \includegraphics[width=\columnwidth]{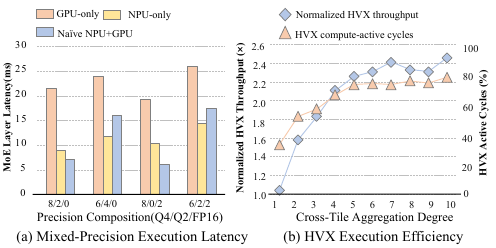}
    \caption{Mixed-precision execution characteristics.}
    \label{fig:mixed_precision_execution}
    \vspace{-4mm}
\end{figure}
\textbf{\textit{Routing Score--Quantization Sensitivity Misalignment.}}
To assess whether routing scores can reliably guide runtime precision assignment in extra-large MoE models, we examine their relationship with expert quantization sensitivity using Qwen3-Next-80B-A3B.
In Layer 18, activated experts are individually quantized to Q4\_0 with other experts unchanged, and their quantization sensitivity is measured using final-logit KL divergence against the FP16 reference model.
Routing scores are normalized across activated experts.

As shown in Fig.~\ref{fig:expert_precision}(a), some experts with low routing scores exhibit high quantization sensitivity.
This occurs because routing scores determine expert output aggregation, but quantization-induced output deviations depend on the interaction between weight quantization errors and MoE input activations.
These deviations can be amplified or attenuated across subsequent layers, resulting in different impacts on the final-logit distribution.
\textbf{\textit{Routing scores cannot accurately characterize expert quantization sensitivity.}}

\textbf{\textit{Critical Dimension--Quantization Sensitivity Correlation.}}
For a quantized expert, weight quantization errors remain fixed, but the resulting output deviations depend on MoE input activations.
We therefore examine the correlation between activation magnitudes in individual MoE input dimensions and final-logit KL divergence.

As shown in the upper part of Fig.~\ref{fig:expert_precision}(b), only a few input dimensions exhibit strong correlations with expert quantization sensitivity.
We further compare normalized activations in the top-32 critical dimensions and random dimensions across tokens ordered by final-logit KL.
As shown in the lower part of Fig.~\ref{fig:expert_precision}(b), critical dimensions exhibit increasing activation magnitudes with higher quantization sensitivity, whereas random dimensions show no clear trend.
Each input dimension contributes to quantization-induced output deviations through its activation and corresponding weight quantization errors.
Since these errors differ across dimensions, activation magnitudes can exhibit different correlations with final-logit KL, consistent with the contrasting trends of critical and random dimensions.
Therefore, \textbf{\textit{activation magnitudes in critical MoE input dimensions exhibit a strong correlation with expert quantization sensitivity.}}

\begin{figure*}[t]
    \centering
    \includegraphics[width=\textwidth]{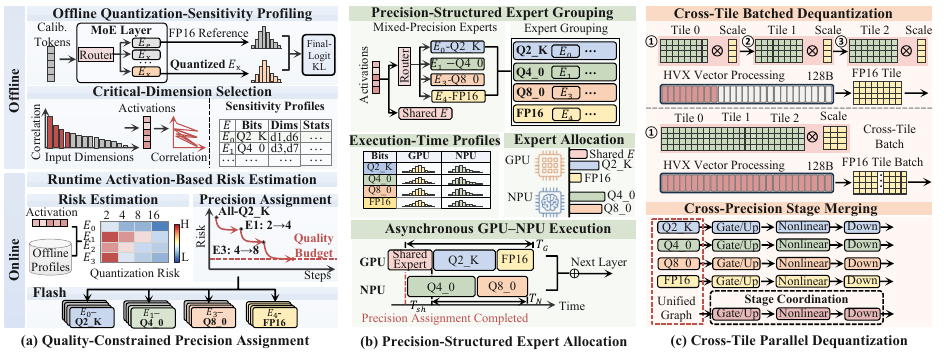}
    \caption{Orchestration techniques in \ours{}.}
    \label{fig:techniques}
    \vspace{-4mm}
\end{figure*}

\subsection{Mixed-Precision Execution Characteristics}

\textbf{\textit{GPU--NPU Workload Imbalance.}}
Activated experts in mixed-precision MoE workloads are assigned one of several bitwidths.
A naive approach allocates activated experts based on estimated computational workloads while accounting for expert bitwidths, with the higher-throughput NPU handling larger workloads than the GPU.
To evaluate this approach, we fix 10 activated experts in Layer 18 and compare GPU-only, NPU-only, and naive GPU--NPU execution across different precision compositions.

As shown in Fig.~\ref{fig:mixed_precision_execution}(a), naive GPU--NPU execution outperforms NPU-only under 8/2/0 and 8/0/2 (Q4/Q2/FP16), but underperforms under 6/4/0 and 6/2/2.
This performance reversal reflects the distinct execution costs of different-bitwidth experts on GPUs and NPUs, which computational workloads cannot fully capture.
As precision compositions change, the same allocation rule can yield imbalanced execution times between the GPU and NPU.
The slower accelerator then determines overall expert computation latency, potentially offsetting the benefits of heterogeneous parallel execution.
Therefore, \textbf{\textit{naive expert allocation based on computational workloads may fail to achieve balanced GPU--NPU execution.}}

\textbf{\textit{NPU Underutilization.}}
During execution, each weight tile is unpacked and scaled independently, which limits the amount of data HVX can process at once.
To explore opportunities for improving dequantization efficiency, we investigate the potential of cross-tile processing in weight dequantization.
We define the \textit{cross-tile aggregation degree} as the number of weight tiles processed together with the same bitwidth and dequantization configurations.
With the total weight size fixed, we increase this degree and measure normalized HVX throughput and the proportion of HVX compute-active cycles.

As shown in Fig.~\ref{fig:mixed_precision_execution}(b), both metrics increase substantially with cross-tile aggregation degree, with diminishing gains as preparation overhead is amortized and HVX utilization approaches saturation.
Cross-tile aggregation allows unpacking and scaling configurations to be reused across multiple tiles, reducing repeated setup and control overhead at tile boundaries.
As this overhead is amortized and HVX utilization approaches saturation, further aggregation yields smaller improvements.
However, diverse bitwidths and dequantization configurations in mixed-precision MoE workloads make it difficult to group compatible weight tiles for cross-tile execution.
Therefore, \textbf{\textit{efficient cross-tile dequantization requires organizing compatible weight tiles to improve HVX utilization.}}

\section{Methods}
\subsection{Quality-Constrained Precision Assignment}

\noindent\textbf{\textit{Offline Quantization-Sensitivity Profiling.}}
During offline calibration, \ours{} collects MoE input hidden states $h_{l,t}\in\mathbb{R}^{d}$ for activated experts across different layers.
\ours{} supports four expert precision formats, including Q2\_K, Q4\_0, Q8\_0, and FP16, with nominal bitwidths $b\in\{2,4,8,16\}$.
Using FP16 as the reference, we measure each activated expert's quantization sensitivity at different bitwidths while keeping other experts' precisions and routing decisions fixed:
\begin{equation}
D_{l,e}^{(b)}(t)
=
D_{\mathrm{KL}}
\left(
p_t^{\mathrm{ref}}
\parallel
p_t^{(l,e,b)}
\right),
\end{equation}
where $p_t^{\mathrm{ref}}$ and $p_t^{(l,e,b)}$ denote the output distributions of the FP16 model and the model with expert $e$ at bitwidth $b$, respectively.

The correlation between activation magnitudes in individual MoE input dimensions and quantization sensitivity is then analyzed.
For each expert and quantized bitwidth, we compute the Spearman correlation across calibration tokens:
\begin{equation}
\rho_{l,e,j}^{(b)}
=
\operatorname{Spearman}_{t\in\mathcal{T}_{l,e}}
\left(
|h_{l,t,j}|,
D_{l,e}^{(b)}(t)
\right),
\end{equation}
where $\mathcal{T}_{l,e}$ denotes the calibration tokens activating expert $e$ in layer $l$.
Based on these correlations, we select $K$ input dimensions with the largest absolute correlation coefficients to form the critical-dimension set $\mathcal{C}_{l,e}^{(b)}$.
The correlation coefficients of the selected dimensions are normalized to obtain dimension-specific correlation weights:
\begin{equation}
w_{l,e,j}^{(b)}
=
\frac{\rho_{l,e,j}^{(b)}}
{\sum_{k\in\mathcal{C}_{l,e}^{(b)}}|\rho_{l,e,k}^{(b)}|},
\quad j\in\mathcal{C}_{l,e}^{(b)}.
\end{equation}
The selected critical-dimension indices and their normalized correlation weights are stored as offline quantization-sensitivity profiles for each expert and quantized bitwidth.

\noindent\textbf{\textit{Runtime Activation-Based Risk Estimation.}}
For each activated expert and quantized bitwidth, \ours{} computes an activation assessment score using the current token's critical-dimension activations and offline correlation weights:
\begin{equation}
A_{l,e}^{(b)}(t)
=
\sum_{j\in\mathcal{C}_{l,e}^{(b)}}
w_{l,e,j}^{(b)}
\cdot
\operatorname{clip}
\left(
\frac{|h_{l,t,j}|-\mu_{l,e,j}}
{\sigma_{l,e,j}+\epsilon},
-\tau,\tau
\right),
\end{equation}
where $\mu_{l,e,j}$ and $\sigma_{l,e,j}$ denote the offline mean and standard deviation of activation magnitudes for calibration tokens activating expert $e$ in layer $l$, respectively; $\epsilon$ ensures numerical stability.
The clipping threshold $\tau$ limits the influence of activation outliers.

At runtime, \ours{} estimates quantization risk by scaling the offline mean quantization sensitivity with the current activation assessment score:
\begin{equation}
r_{l,e}^{(b)}(t)
=
\bar D_{l,e}^{(b)}
\exp\left(A_{l,e}^{(b)}(t)\right),
\end{equation}
where $\bar D_{l,e}^{(b)}$ denotes the mean quantization sensitivity measured over calibration tokens activating expert $e$ in layer $l$ at bitwidth $b$.
For FP16, the estimated quantization risk is set to zero.

\noindent\textbf{\textit{Quality-Constrained Precision Assignment.}}
Based on the estimated quantization risks, \ours{} dynamically selects expert bitwidths under a model quality budget.
For the activated expert set $\mathcal{E}_l(t)$ in layer $l$, we formulate precision assignment as:

\begin{equation}
\min_{\{b_{l,e}\in\{2,4,8,16\}\}}
\sum_e b_{l,e}
\quad \mathrm{s.t.} \quad
\sum_e r_{l,e}^{(b_{l,e})}(t)\leq\mathcal{B}_l.
\end{equation}

where $\mathcal{B}_l$ denotes the calibrated quality budget for layer $l$, and both summations are over the activated expert set $\mathcal{E}_l(t)$.
To determine $\mathcal{B}_l$, we analyze the relationship between aggregated quantization risks and final-logit KL divergence across representative precision assignments on calibration tokens.
We then translate a predefined KL tolerance into a layer-specific risk budget based on this relationship.

For each token at a MoE layer, \ours{} initializes all activated experts to Q2\_K.
If the total estimated risk exceeds the layer-specific quality budget, \ours{} iteratively upgrades the expert whose next higher precision offers the largest risk reduction per added bitwidth.
The process stops once the quality budget is satisfied, enabling dynamic precision assignment while favoring lower bitwidths.

\subsection{Precision-Structured Expert Allocation}

\noindent\textbf{\textit{Precision-Structured Expert Grouping.}}
Unlike routed experts, shared experts require neither gating nor precision assignment.
\ours{} therefore dispatches them asynchronously to the GPU once the MoE input becomes available, overlapping their execution with gating computation.
Once gating and precision assignment are complete, \ours{} organizes routed experts into same-precision groups based on their bitwidths.
These groups serve as the basic organizational units for subsequent GPU--NPU expert allocation.

\noindent\textbf{\textit{Execution-Aware Expert Allocation.}}
Based on the same-precision expert groups, \ours{} dynamically allocates experts across the GPU and NPU using device-specific execution-time estimates.
To estimate execution costs, \ours{} performs offline profiling of expert-group execution times on GPU and NPU across bitwidths and group sizes, covering weight loading, format preparation, computation, and data transfer.
At runtime, \ours{} determines the GPU--NPU split for each precision group based on the profiled execution costs and the current token's precision composition.

For each precision $p$, let $n_p$ denote the number of activated routed experts in the group.
Among them, $x_p$ experts are allocated to the NPU and the remaining $n_p-x_p$ to the GPU.
Let $c_{d,p}(n)$ denote the estimated execution time of $n$ experts with precision $p$ on device $d\in\{\mathrm{G},\mathrm{N}\}$, with $c_{d,p}(0)=0$.
At the allocation point, \ours{} estimates GPU and NPU completion times, accounting for the remaining shared-expert execution on the GPU:
\begin{equation}
\begin{aligned}
T_{\mathrm G}(\mathbf{x}) &= T_{\mathrm{sh}} + \sum_{p\in\mathcal P} c_{\mathrm G,p}(n_p-x_p),\\
T_{\mathrm N}(\mathbf{x}) &= \sum_{p\in\mathcal P} c_{\mathrm N,p}(x_p),
\end{aligned}
\end{equation}
where $\mathcal P$ denotes the set of active expert precisions, $\mathbf{x}=\{x_p\}_{p\in\mathcal P}$ denotes the NPU allocation vector, and $T_{\mathrm{sh}}$ represents the remaining GPU execution time of shared experts at this point.
To minimize expert execution latency, \ours{} formulates the allocation problem as:
\begin{equation}
\begin{aligned}
\min_{\mathbf{x}} \quad
& \max\left(T_{\mathrm G}(\mathbf{x}),T_{\mathrm N}(\mathbf{x})\right)\\
\mathrm{s.t.} \quad
& x_p\in\{0,1,\ldots,n_p\},\quad \forall p\in\mathcal P.
\end{aligned}
\end{equation}
\ours{} enumerates feasible GPU--NPU allocations and selects the one minimizing the predicted maximum completion time.

\noindent\textbf{\textit{Asynchronous GPU--NPU Execution.}}
After allocation, \ours{} asynchronously dispatches the assigned routed-expert workloads to the GPU and NPU for concurrent execution.
Upon completion of both workloads, \ours{} aggregates routed-expert outputs using their original routing weights and incorporates shared-expert outputs according to the model's computation semantics.

\subsection{Cross-Tile Parallel Dequantization}
\noindent\textbf{\textit{Cross-Tile Batched Dequantization.}}
For each quantized same-precision group formed in Section~4.2, \ours{} organizes the weights of its NPU-assigned experts into dequantization tiles.
It then aggregates $m$ tiles with compatible unpacking and scaling configurations into a cross-tile batch for batched dequantization.

In conventional tile-wise execution, each quantized weight tile requires separate scale metadata preparation, weight unpacking, and dequantization before matrix--vector computation.
For each output block $o$ of expert $e$ with precision $p$, the tile-wise partial results are accumulated along the input dimension:
\begin{equation}
\mathbf{y}_{e,o}
=
\sum_{i=1}^{R_{e,o}}
\operatorname{Dequant}_{p}
(\mathbf{Q}_{e,o,i},\mathbf{s}_{e,o,i})
\mathbf{a}_{e,i},
\end{equation}
where $R_{e,o}$ denotes the number of input-dimension tiles, while $\mathbf{Q}_{e,o,i}$, $\mathbf{s}_{e,o,i}$, and $\mathbf{a}_{e,i}$ represent the compressed weights, scale metadata, and corresponding input activation slice of tile $i$, respectively.
The operator $\operatorname{Dequant}_{p}$ performs weight unpacking and numerical reconstruction for precision $p$.
In contrast, \ours{} packs the compressed weights and per-tile scale metadata of each cross-tile batch for processing by a single HVX kernel.
The kernel reuses unpacking and scaling configurations to perform vectorized dequantization across tiles, producing FP16 weights for subsequent HMX matrix computation.
For each tile $(e,o,i)$, let $\mathcal I_{e,o,i}$ denote its cross-tile batch with $m$ tile slots, padding incomplete batches as needed.
The original expert computation is preserved as:
\begin{equation}
\begin{aligned}
\mathbf{y}_{e,o}
=\sum_{i=1}^{R_{e,o}}
\Big[
\operatorname{Dequant}_{p}
\big(\mathbf{Q}_{\mathcal{I}_{e,o,i}},
\mathbf{S}_{\mathcal{I}_{e,o,i}}\big)
\Big]_{e,o,i}
\mathbf{a}_{e,i},
\end{aligned}
\end{equation}
where $\mathbf Q_{\mathcal I}$ and $\mathbf S_{\mathcal I}$ denote the batched compressed weights and per-tile scale metadata, respectively.
For batched inputs, $\operatorname{Dequant}_{p}$ reconstructs multiple weight tiles within one HVX kernel invocation, while $[\cdot]_{e,o,i}$ extracts the FP16 weight tile.
By reusing configurations across compatible tiles, \ours{} amortizes setup and execution control overhead, improving HVX vector utilization and dequantization throughput without changing the original matrix computation.

\noindent\textbf{\textit{Cross-Precision Stage Merging.}}
Beyond cross-tile batched dequantization, separate graph executions for each precision group and computation stage incur repeated submission and synchronization overhead.
To reduce this overhead, \ours{} merges Gate/Up computations across precision groups into one NPU execution graph and Down computations into another.
The two stage-specific graphs preserve the nonlinear activations, while each precision group retains its original quantization format and dequantization operations.

\begin{table}[!t]
    \centering
    \footnotesize
    \renewcommand{\arraystretch}{1.18}
    \setlength{\tabcolsep}{2pt}

    \caption{Configurations of MoE models.}
    \label{tab:models}
    \begin{tabular*}{\columnwidth}{@{\extracolsep{\fill}}lccc@{}}
        \hline
        \textbf{Model} &
        \textbf{Layers} &
        \textbf{Experts} &
        \textbf{Routing} \\
        \hline
        Qwen3-Next-80B-A3B & 48 & 512 & Top-10 \\
        Kimi-Linear-48B-A3B & 26 & 256 & Sigmoid Top-8 \\
        Qwen3.6-35B-A3B & 40 & 256 & Top-8 \\
        \hline
    \end{tabular*}

    \par\vspace{8pt}

    \caption{Output distribution deviation.}
    \label{tab:precision_analysis}
    \begin{tabular*}{\columnwidth}{@{\extracolsep{\fill}}cccc@{}}
        \hline
        \shortstack[c]{\textbf{Avg. Expert}\\\textbf{Bitwidth}} &
        \shortstack[c]{\textbf{Frequency-Based}\\\textbf{Offline}} &
        \shortstack[c]{\textbf{Routing-Based}\\\textbf{Online}} &
        \raisebox{5pt}{\textbf{MixMoMo} $\downarrow$} \\[-1.3pt]
        \hline
        2.5 & 0.672 & 0.265 & \textbf{0.194} \\
        3.0 & 0.634 & 0.153 & \textbf{0.105} \\
        3.5 & 0.601 & 0.132 & \textbf{0.084} \\
        4.0 & 0.653 & 0.095 & \textbf{0.033} \\
        4.5 & 0.667 & 0.091 & \textbf{0.028} \\
        5.0 & 0.593 & 0.088 & \textbf{0.015} \\
        6.0 & 0.595 & 0.063 & \textbf{0.009} \\
        \hline
    \end{tabular*}

    \par\vspace{8pt}

    \caption{Accuracy Across Mixed-Precision Methods.}
    \label{tab:accuracy}
    \begin{tabular*}{\columnwidth}{@{\extracolsep{\fill}}lccc@{}}
        \hline
        \raisebox{5pt}{\textbf{Method}} &
        \shortstack[c]{\textbf{GSM8K}\\\textbf{Acc. (\%) $\uparrow$}} &
        \shortstack[c]{\textbf{HumanEval}\\\textbf{Pass@1 (\%) $\uparrow$}} &
        \shortstack[c]{\textbf{MMLU-Pro}\\\textbf{Acc. (\%) $\uparrow$}} \\
        \hline
        FP16          & 84.53 & 92.12 & 76.35 \\
        Uniform Q4\_0 & 80.46 & 87.94 & 71.48 \\
        EdgeMoE       & 81.45 & 88.43 & 72.65 \\
        MoE-APEX      & 82.23 & 90.93 & 72.23 \\
        DyMoE         & 82.98 & 90.72 & 73.46 \\
        MxMoE         & 83.19 & 91.06 & 74.92 \\
        MixMoMo       & \textbf{84.48} & \textbf{92.00} & \textbf{75.87} \\
        \hline
    \end{tabular*}
\end{table}

\section{Implementation}
\ours{} is implemented by extending llama.cpp~\cite{gerganov2023llamacpp} with approximately 8,000 lines of C/C++ code across its heterogeneous inference runtime, OpenCL backend, and Qualcomm Hexagon HTP backend.
To reduce storage overhead, \ours{} uses a Matryoshka-style shared representation~\cite{nair2025matquant} for mixed-precision expert weights and loads weights at the required bitwidth on demand.
The offline quantization-sensitivity profiles described in Section~4.1 occupy approximately 18 MiB of DRAM, with 16 critical input dimensions per expert and quantized bitwidth.
Since routed experts within each evaluated model share identical matrix dimensions across MoE layers~\cite{qwen2025qwen3next,kimiteam2025kimilinear,qwen2026qwen36}, Section~4.2 reuses precomputed model execution-time profiles across layers.
GPU and NPU execution times are estimated based on expert bitwidth and group size, and these profiles occupy approximately 0.24 KiB of DRAM.
The batch parallelism in Section~4.3 is set to 6 weight tiles because processing more tiles in parallel provides only marginal improvements in HVX throughput.

\section{Experiments}
\subsection{Experimental Setup}

\begin{figure*}[t]
    \centering
    \includegraphics[width=\textwidth]{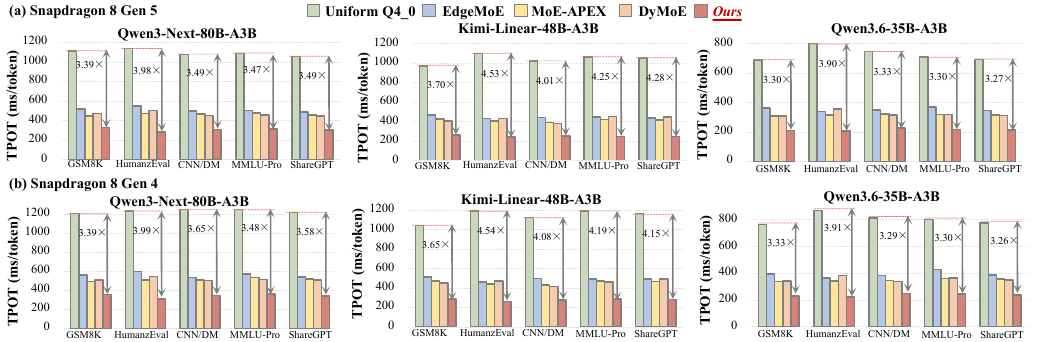}
    \vspace{-4mm}
    \caption{Decode performance.}
    \label{fig:decode}
    \vspace{-4mm}
\end{figure*}

\noindent\textbf{\textit{Hardware Platform.}}
\ours{} is evaluated on the OnePlus 15 with Snapdragon 8 Elite Gen 5 (8Gen5) and the OnePlus 13 with Snapdragon 8 Elite (8Gen4).
The tested OnePlus 15 features 16 GB DRAM and 256 GB UFS 4.1 flash storage, while the tested OnePlus 13 features 24 GB DRAM and 1 TB UFS 4.0 flash storage.

\noindent\textbf{\textit{Datasets.}}
We evaluate model quality and inference performance on five datasets: GSM8K, HumanEval, CNN/DailyMail, MMLU-Pro, and ShareGPT.
These datasets cover mathematical reasoning, code generation, news summarization, knowledge-intensive question answering, and conversational dialogue, respectively.
The maximum number of generated tokens is set to 256 for all datasets.

\noindent\textbf{\textit{Models.}}
Three representative MoE models from the Qwen and Kimi families are selected for evaluation: Qwen3-Next-80B-A3B~\cite{qwen2025qwen3next}, Kimi-Linear-48B-A3B~\cite{kimiteam2025kimilinear}, and Qwen3.6-35B-A3B~\cite{qwen2026qwen36}.
As detailed in Table~\ref{tab:models}, the evaluated models range from 35B to 80B total parameters, with diverse expert counts and routing strategies.

\noindent\textbf{\textit{Baselines.}}
To evaluate the inference performance of \ours{}, we compare it against a uniform quantization baseline and three representative MoE inference systems for mobile and edge platforms.
(i) \textbf{\textit{Uniform Q4\_0.}} All expert weights are uniformly quantized to Q4\_0.
(ii) \textbf{\textit{EdgeMoE}}~\cite{yi2025edgemoe}. It assigns expert bitwidths based on offline quantization sensitivity and employs expert prediction and preloading to reduce weight loading overhead.
(iii) \textbf{\textit{MoE-APEX}}~\cite{tang2026moeapex}. It selects cache-miss expert bitwidths based on accumulated routing scores and optimizes weight loading through layer-level prefetching and sequence-level caching.
(iv) \textbf{\textit{DyMoE}}~\cite{huang2026dymoe}. It dynamically assigns expert bitwidths based on routing weights and layer-depth sensitivity and employs look-ahead prefetching to reduce loading overhead.
Since these baselines combine different precision assignment and weight management strategies, we compare their overall inference performance rather than precision assignment alone.
The contributions of \ours{}'s individual runtime optimizations are further analyzed through an incremental ablation study.

\noindent\textbf{\textit{Metrics.}}
We evaluate model quality using final-logit KL divergence and task accuracy.
For precision assignment, we compare final-logit KL divergence under matched average expert bitwidth budgets.
For inference performance, all mixed-precision methods use an average expert bitwidth of 4 bits, with the quality budget of \ours{} adjusted to meet this target.
We report Time Per Output Token (TPOT, ms/token) and end-to-end latency to evaluate inference performance.

\subsection{Evaluation Results}


\textbf{\textit{Precision Assignment Effectiveness.}}
We measure final-logit KL divergence against the full-precision reference across average expert bitwidths to quantify output deviations.
On Qwen3-Next-80B-A3B, we compare frequency-based, routing-based~\cite{tang2026moeapex,kim2026dpamoe}, and \ours{} assignment.
Table~\ref{tab:precision_analysis} shows that \ours{} achieves lower KL divergence than both baselines at comparable bitwidths.
Near-uniform expert activation in extra-large MoEs undermines frequency-based precision assignment.
Routing scores do not accurately reflect expert quantization effects on model outputs.
In contrast, \ours{} combines offline quantization-impact profiles with current-token activations for more accurate quantization risk estimation.


\noindent\textbf{\textit{Accuracy.}}
Model accuracy is assessed on Qwen3-Next-80B-A3B using GSM8K and MMLU-Pro accuracy and HumanEval Pass@1.
We further include MxMoE~\cite{duanmu2025mxmoe} as a dedicated mixed-precision MoE inference baseline alongside FP16 and the existing methods.
Table~\ref{tab:accuracy} shows that \ours{} incurs minimal accuracy degradation compared with FP16.
It outperforms all quantized baselines across the three benchmarks.
These results further demonstrate that the quality budget in \ours{} limits quantization-induced accuracy degradation.


\textbf{\textit{Decode Performance.}}
We compare the TPOT of \ours{} against Uniform Q4\_0 and existing mixed-precision baselines across three MoE models and five benchmarks on both Snapdragon platforms.
As shown in Fig.~\ref{fig:decode}, \ours{} achieves the lowest TPOT across all models and benchmarks on both platforms.
The average speedups over Uniform Q4\_0 are 3.61$\times$, 4.06$\times$, and 3.49$\times$ for Qwen3-Next-80B-A3B, Kimi-Linear-48B-A3B, and Qwen3.6-35B-A3B, respectively.
Compared with the strongest existing baseline, \ours{} further improves decoding throughput by 1.50$\times$, 1.61$\times$, and 1.48$\times$, respectively.
Precision-Structured Expert Allocation fully utilizes GPU compute resources to complement NPU execution, while HVX-Parallel Dequantization improves NPU execution efficiency, thereby reducing expert computation latency and overall TPOT.

\begin{figure}[t]
    \centering
    \includegraphics[width=\columnwidth]{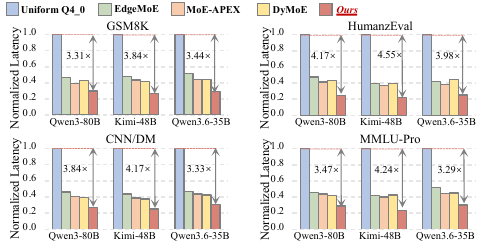}
    \caption{End-to-end Latency.}
    \label{fig:e2e_latency}
    \vspace{-4mm}
\end{figure}


\textbf{\textit{End-to-End Latency.}}
Due to substantial execution time differences across benchmarks, end-to-end latency is normalized against Uniform Q4\_0 for clearer comparisons across three MoE models and four benchmarks.
As shown in Fig.~\ref{fig:e2e_latency}, \ours{} achieves the lowest end-to-end latency across all models and benchmarks.
Compared with the strongest baseline for each model, \ours{} reduces average end-to-end latency by 32.52\%, 36.55\%, and 30.49\% for Qwen3-Next-80B-A3B, Kimi-Linear-48B-A3B, and Qwen3.6-35B-A3B, respectively.
These results demonstrate that \ours{} consistently reduces end-to-end latency despite differences in benchmark execution times.

\textbf{\textit{Ablation Study.}}
\begin{table}[t]
    \centering
    \caption{Incremental Ablation of MixMoMo.}
    \label{tab:ablation}
    \footnotesize
    \renewcommand{\arraystretch}{1.18}
    \setlength{\tabcolsep}{2pt}
    \begin{tabular*}{\columnwidth}{@{\extracolsep{\fill}}lcccc@{}}
        \hline
        \raisebox{5pt}{\textbf{Config.}} &
        \shortstack[c]{\textbf{GSM8K}\\\textbf{TPOT (ms) $\downarrow$}} &
        \shortstack[c]{\textbf{HumanEval}\\\textbf{TPOT (ms) $\downarrow$}} &
        \shortstack[c]{\textbf{CNN/DM}\\\textbf{TPOT (ms) $\downarrow$}} &
        \shortstack[c]{\textbf{Avg.}\\\textbf{Speedup $\uparrow$}} \\
        \hline
        Uniform Q4\_0 & 1111.11 & 1136.36 & 1075.27 & 1.00$\times$ \\
        + QCPA        & 1123.59 & 1062.73 & 1091.35 & 1.01$\times$ \\
        + PSEA        & 498.26  & 452.96  & 482.11  & 2.32$\times$ \\
        + CTPD         & \textbf{327.87} & \textbf{284.90} & \textbf{308.64} & \textbf{3.61$\times$} \\
        \hline
    \end{tabular*}
    \par\vspace{4pt}
    \parbox{\columnwidth}{\scriptsize
        \textit{Note:} QCPA: Quality-Constrained Precision Assignment;
        PSEA: Precision-Structured Expert Allocation;
        CTPD: Cross-Tile Parallel Dequantization.
    }
    \vspace{-4mm}
\end{table}
An incremental ablation study is conducted on Qwen3-Next-80B-A3B with Uniform Q4\_0 as the baseline.
Quality-Constrained Precision Assignment, Precision-Structured Expert Allocation, and Cross-Tile Parallel Dequantization are sequentially enabled to quantify their TPOT contributions.
As shown in Table~\ref{tab:ablation}, Quality-Constrained Precision Assignment introduces only minor TPOT variations compared with Uniform Q4\_0.
These variations arise from dynamic expert bitwidth assignment, which changes the average expert precision and corresponding loading and computation costs.
While Quality-Constrained Precision Assignment controls quantization-induced quality degradation, Precision-Structured Expert Allocation reduces TPOT by 56.27\% over precision assignment alone through precision-aware expert grouping and execution-aware GPU--NPU co-execution.
Cross-Tile Parallel Dequantization further reduces TPOT by 35.72\% through improved HVX vector utilization and reduced weight preparation and synchronization overhead.
These results highlight the complementary roles of precision assignment in preserving model quality and heterogeneous execution optimizations in accelerating inference.
Together, these mechanisms enable \ours{} to achieve quality-constrained, low-latency mixed-precision inference of extra-large MoE models on heterogeneous mobile SoCs.

\section{Conclusion}
Efficient mixed-precision inference of extra-large MoE models on heterogeneous mobile SoCs is challenged by unreliable expert precision assignment, imbalanced GPU--NPU execution, and inefficient NPU dequantization.
To address these challenges, we present \ours{}, which combines activation-based quality-constrained precision assignment, precision-structured expert allocation, and cross-tile parallel dequantization.
Evaluations across three MoE models and five benchmarks on two smartphones show that \ours{} maintains near-FP16 model quality while achieving average decoding speedups of 3.72$\times$ over uniform Q4\_0 and 1.53$\times$ over the strongest baseline.

\bibliographystyle{IEEEtran}
\bibliography{references}
\end{document}